\documentclass[prd,twocolumn,amsmath,amssymb,aps,superscriptaddress,groupedaddress,nofootinbib,showpacs]{revtex4-2}
\usepackage{caption}

\usepackage{graphicx,float}
\usepackage{subcaption}
\usepackage[usenames, dvipsnames]{color}
\usepackage{tabularx}
\usepackage{siunitx,booktabs}
\usepackage{xcolor}
\usepackage{ragged2e} 
\usepackage{caption}
\usepackage[
verbose,
colorlinks=true,
naturalnames=true,
linkcolor=blue,
citecolor=blue,
]{hyperref}

\preprint{APS/123-QED}
\begin{document}
	
	\title{
		Global Chlorophyll-\textit{a} Retrieval algorithm from Sentinel 2 Using Residual Deep Learning and Novel Machine Learning  Water Classification 
	}
	\author{
		Yotam Sherf~$^{1}$, Bar Efrati, Gabriel Rozman, and Moshe Harel\\
		\small{\textit{BlueGreen Water Research Center, HaMiktsot Blvd. 16, Modi'in Makabim-Re'ut, 7178096, Israel}}}

	\begin{abstract}
		We present the Global Water Classifier (GWC), a supervised, geospatially extensive Machine Learning (ML) classifier trained on Sen2Cor corrected Sentinel-2 surface reflectance data. Using nearly 100 globally distributed inland water bodies, GWC distinguishes water across  Chlorophyll-\textit{a} (Chl\textit{a}) levels from non-water spectra (clouds, sun glint, snow, ice, aquatic vegetation, land and sediments) and shows geographically stable performance.
		Building on this foundation model, we perform Chl\textit{a} retrieval based on a matchup Sentinel-2 reflectance data with the United States Geological Survey (USGS) AquaMatch in-situ dataset, covering diverse geographical and hydrological conditions. 
		We train an XGBoost regressor on $13626$ matchup points. The positive labeled scenes by the GWC consistently outperform the negatives and produce more accurate Chl\textit{a} retrieval values, which confirms the  classifier’s advantage in reducing various interferences. 
		Next, residual analysis of the regression predictions revealed structured errors, motivating a residual CNN (RCNN) correction stage. We add a CNN residual stage trained on normalized residuals, which yield substantial improvement. Our algorithm was tested on 867 water bodies with over 2,000 predictions and Chl\textit{a} values up to 1000~mg$/m^{3}$, achieving $R^2 = 0.79$, MAE = 13.52~mg$/m^{3}$, and slope = 0.91, demonstrating robust, scalable, and globally transferable performance without additional tuning.
		
	\end{abstract}
	\maketitle
	Harmful algal blooms\footnotetext{Corresponding author: ysherf@bluegreenwatertech.com}
	 (HABs) have become increasingly prevalent across inland and coastal waters, threatening drinking water quality, producing toxins and greenhouse gasses, and disrupting aquatic ecosystems with mounting economic and public health costs \cite{gobler2020climate,paerl2016controlling,kaplan_harel_2025_bloomGHG}. Their growing frequency and intensity underscore the urgent need for continuous, spatially extensive monitoring. Traditional field sampling, while accurate, is costly and geographically limited, prompting the rapid expansion of satellite-based monitoring as an indispensable complement to in situ observations.
	Satellite remote sensing enables global-scale HAB surveillance with frequent revisit times and high spatial resolution. Instruments such as Sentinel-2’s MultiSpectral Imager (MSI) provide imagery at 10–20 m resolution every five days, making them particularly valuable for small and optically complex inland waters \cite{ioccg2006report5,mishra2013phycocyanin,llodra2023review}. 
	
	Yet, accurate Chl\textit{a} retrieval, the principal proxy for phytoplankton biomass, remains challenging, as observations are prone to natural and instrumental perturbations that can substantially alter the expected optical response of Chl\textit{a} \cite{pahlevan2019seamless,moses2012evaluation,llodra2023review}. Ideally, we would expect a predictable linear (or at least near linear) relationship of Chl\textit{a} to the spectral imprint. However, in practice, the signal deviates due to non-algal particles such as turbidity, colored dissolved organic matter (CDOM), sediments, and others, which modify the optical signature. At high Chl\textit{a} levels (often referred as scum), optical saturation and multiple non-linear scattering further distort the spectrum  \cite{moses2012evaluation, pahlevan2019seamless}.  In addition, imaging artifacts such as sun glint, cloud shadows, snow or ice, and adjacency effects frequently degrade data quality. Collectively, these optical and atmospheric perturbations introduce nonlinear and nonstationary distortions that complicate model transferability across lakes, seasons, and sensors \cite{he2022deep,llodra2023review}.

	Despite these challenges, empirical algorithms remain widely used in operational monitoring systems due to their transparency, computational simplicity, and compatibility with long-term archives \cite{matthews2014eutrophication,lobo2021algaemap}. Over the past two decades, diverse strategies have been proposed to improve Chl\textit{a} estimation \cite{moses2012evaluation, lobo2021algaemap, pahlevan2019seamless, gilerson2010retrieval}. Index-based and band-ratio algorithms remain the most interpretable and computationally efficient options but often fail under optically variable conditions \cite{moses2012evaluation,lobo2021algaemap}. Semi-analytical inversion models, which couple reflectance with radiative-transfer theory, improve physical interpretability but depend on precise atmospheric correction and reliable prior knowledge of optical properties \cite{zhang2022bayesian}.
	Increasingly, machine-learning (ML) approaches, including Random Forest (RF), gradient-boosted trees, and deep neural networks, are demonstrating strong potential to model nonlinear relationships between reflectance and Chl\textit{a} across diverse optical regimes \cite{he2022deep, joshi2024cross,li2025mlretrieval}. Recent advances in interpretable and hybrid learning architectures have further improved model transparency and physical consistency \cite{zhang2022bayesian,zhong2022interpretable}, enabling more robust generalization across optical conditions. These models have been successfully applied to multisensor datasets (e.g., Sentinel-3 OLCI, MODIS, and MSI) to predict algal biomass and detect bloom dynamics at regional to continental scales \cite{pahlevan2019seamless, he2022deep}. However, domain shift, inconsistent atmospheric corrections, and limited labeled training data remain critical barriers to reliable generalization \cite{wynne2010detecting, kudela2015habsummary,gilerson2010retrieval,yacobi2011phytoplankton,dev2022a,dev2022b,dasilva2024chlorophyll}, as recent multi-lake studies have demonstrated steep performance declines outside the training domain \cite{dasilva2024chlorophyll}. Additionally, many of these prior studies optimized for a single lake or region, which is not applicable across diverse inland waters without site-specific tuning \cite{joshi2024cross, dasilva2024chlorophyll}.
	Recent operational systems such as AlgaeMAp for Latin American inland waters demonstrate the feasibility of near-real-time bloom monitoring at large scales, but still rely primarily on empirical indices and thresholding rather than advanced data-driven frameworks \cite{lobo2021algaemap}. At the same time, emerging deep-learning methods that integrate physical constraints and spectral–spatial fusion have shown notable improvements in capturing complex optical variability and cross-sensor consistency \cite{he2022deep,zhang2022bayesian}. Our framework leverages this paradigm by coupling global water classification with interpretable residual learning, integrating physical understanding with data-driven inference \cite{zhong2022interpretable,zhang2022bayesian}. These developments collectively emphasize the need for robust, generalizable frameworks that unify data cleaning, feature learning, and predictive modeling across heterogeneous aquatic environments.
	To address these challenges, this study introduces a three-stage ML framework for global Chl\textit{a} mapping from Sentinel-2 MSI imagery. The first stage employs a Global Water Classifier (GWC) trained on globally distributed spectra to probabilistically retain reliable water pixels and remove environmental artefacts such as clouds, ice, aquatic vegetation, and sediment. The second stage applies an XGBoost regression model trained on 13,626 georeferenced in situ Chl\textit{a} samples from the USGS AquaMatch dataset \cite{brousil2024_aquamatch} to learn nonlinear spectral relationships. Finally, a Residual Convolutional Neural Network (RCNN) refines predictions by modeling structured residual errors between observed and predicted Chl\textit{a}, effectively reducing bias while preserving cross-lake generalization. This residual-learning design conceptually parallels recent deep fusion networks that blend physical and data-driven principles for improved optical retrievals \cite{he2022deep,zhang2022bayesian} but extends their application to the residual domain. Together, these stages deliver a globally scalable, physically informed, and data-driven solution for robust Chl\textit{a} estimation across diverse inland waters.
	The novelty of our framework lies in its integrated three-stage workflow. Stage 1 (GWC) acquires analyzable water spectra by removing optical artifacts and isolating true bloom signatures, providing clean inputs for downstream modeling. Stage 2 (XGBoost) learns interaction-rich relationships among bands and indices using ~5,000 AquaMatch in situ points, yielding accurate base Chl\textit{a} predictions consistent with evidence that tree-based learners perform robustly across sensors and optical regimes \cite{joshi2024cross,he2022deep,li2025mlretrieval}. Stage 3 (residual 1-D CNN) identifies and corrects structured residuals across Chl\textit{a} regimes, improving accuracy and preserving cross-lake generalization. The result is a scalable pipeline suitable for diverse geographic regions and environmental conditions, reducing the need for site-specific tuning or auxiliary field measurements beyond Chl\textit{a}.
	The paper is organized as follows. Section 2 describes the datasets, matchup criteria, and preprocessing (Sentinel-2 surface reflectance, indices, and in situ records). Section 3 details the GWC design and training. Section 4 presents the base Chl\textit{a} regression model. Section 5 introduces the RCNN correction and training regimes. Section 6 presents results, including linear and log-space metrics. Section 7 discusses limitations, particularly low-Chl\textit{a} sensitivity and atmospheric-correction artifacts, and Section 8 demonstrates pipeline performance and operational implications.

	\section{Global Water Classifier}

	In this section, we describe the framework developed for the GWC.

	The primary motivation behind this approach is addressing the limitations of traditional hard-coded classification methods, where non-water pixels are masked based on threshold values of spectral indices. These methods tend to be error-prone and lack generalizability. For example, the commonly used Normalized Difference Water Index (NDWI) typically yields positive values for water and negative for land, however, during training, we frequently observed cases where water spectra with varying bloom intensities exhibited negative NDWI value. Similar threshold-based inaccuracies were noted with weak sun glints, sediments, and aquatic vegetation, which were often mistaken for dense algal blooms. Moreover, removing algal bloom spectral mimickers such as aquatic vegetation and certain sediment types is essential, as these can exhibit similar spectral characteristics to blooms under specific conditions and wavelength regions. Some spectral deviations are natural and ecological, reflecting the inherent diversity in water composition and environmental conditions. However, other optical interferences stem from the inaccurate atmospheric correction (Sen2Cor), which exhibits inconsistent performance across different geographical regions. Additionally, noise inherent to Sentinel-2 sensors frequently introduces small-scale spectral variability, further complicating accurate classification. Consequently, there is a clear need for a more flexible, and scalable approach that can isolate true algal bloom signature from limiting spectral interferences. In this study, we address this challenge by developing a globally diverse dataset encompassing a wide range of environmental conditions and allowing identification of water pixels in various bloom conditions worldwide.

	\begin{figure*}[!htbp]
		\centering
		\includegraphics[width=0.75\textwidth]{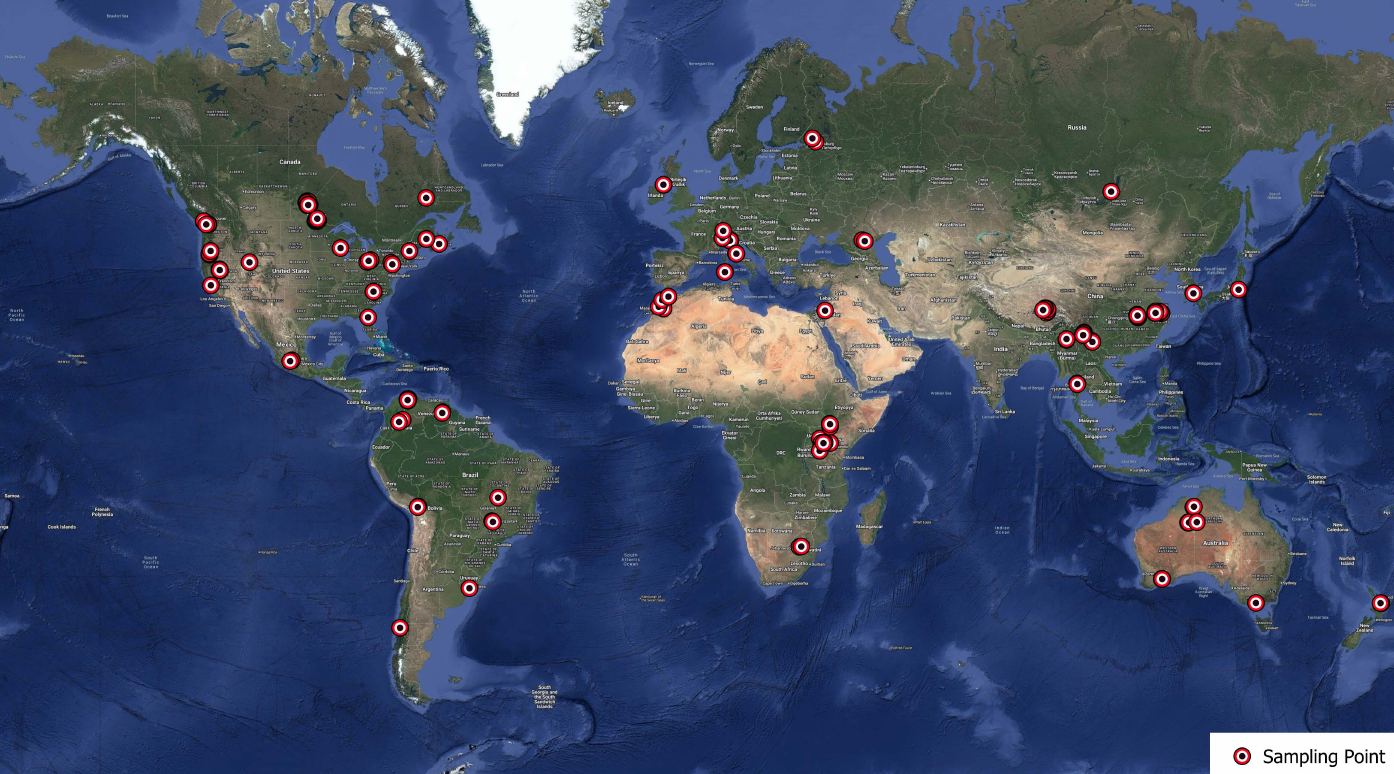}
		\captionsetup{width=\textwidth}
		\caption{\small{The map illustrates the global classification areas, highlighting the broad geographical spread. Classification was carried out across nearly 100 different water bodies, spanning diverse climatic conditions and geographical regions, with more than 500 classification points in total.}}
		\label{globalmap}
	\end{figure*}
	
	\section{Framework}

	The training dataset comprises nearly 100 diverse inland water bodies worldwide, including reservoirs, dams, and lakes of various sizes, with a substantial portion derived from the Gloria dataset \cite{Lehmann2022_GLORIA}. The primary guideline during dataset construction was diversity, ensuring representation from various pixel types, including clouds, sun glints, snow , ice, sediments, aquatic vegetation, land, muddy areas, clear water, and varying intensities of algal blooms. A global distribution map of the training dataset is provided in Fig.\ref{globalmap}.

	We utilized the harmonized Sentinel-2 Surface Reflectance (SR) product from Google Earth Engine (GEE), extracting all atmospherically corrected bands using Sen2Cor: 443, 490, 560, 665, 705, 740, 783, 842, 865, 945, 1610, and 2190 nm. In addition, we derived the following spectral indices:
	
	\begin{gather}
		\text{NDWI} = \frac{R_{560} - R_{865}}{R_{560} + R_{865}},  \nonumber\\
		\text{NDVI} = \frac{R_{665} - R_{865}}{R_{665} + R_{865}},  \nonumber\\
		\text{BI}   = \frac{R_{705}}{R_{665}},\label{si12} \\
		\text{NDTI} = \frac{R_{665} - R_{560}}{R_{665} + R_{560}},  \nonumber\\
		\text{FAI}  = R_{865} - \left[ R_{665} + \left(R_{1610} - R_{665}\right) 
		\frac{\lambda_{865} - \lambda_{665}}{\lambda_{1610} - \lambda_{665}} \right]. \nonumber
	\end{gather}

	where $R_{\lambda}$ denotes the surface reflectance at wavelength $\lambda$. The Floating Algae Index (FAI), originally developed for detecting floating algae in oceans, enhances the discrimination between aquatic vegetation and high algal concentrations \cite{Hu2009_FAI}. The Normalized Difference Turbidity Index (NDTI) measures water transparency influenced by inorganic and organic matter and represents a significant source of interference in Chl\textit{a} detection.

	The classification objective was defined as a binary decision: identifying water pixels, particularly those with varying intensities of algal blooms, and distinguishing them from other pixel types.

	Initially, a model was trained to exclude pixels significantly different from water, such as clouds, land, strong sun glints, and ice, effectively removing primary outliers. In the subsequent stage, we employed a dynamic training approach (see section below) and extended the dataset by incorporating water pixels with various spectral properties, such as various bloom intensities, sediments, and aquatic vegetation. To mitigate potential human labelling errors, the pixels spectrum was verified through individual inspection using standard indices such as NDWI, NDVI, BI and NDTI \cite{McFeeters1996,Chen2005}and were further compared to the surrounding pixels for consistency check.

	Overall, we trained a Random Forest (RF) classifier using approximately 500 different spectral signatures from nearly 100 different water bodies displayed in Fig. \ref{globalmap}, where each contains the mean of aggregated adjacent pixels. Model training employed a 70$\%$-30$\%$ train-test split, with hyperparameter tuning conducted via 5-fold cross-validation. The RF hyperparameters were tuned using the grid search method. Optimizing parameters such as the number of estimators, split criteria, tree depth, minimum samples per split and leaf, and maximum features per split. The optimal model configuration is the Gini criterion, unlimited tree depth, square root of features per split, a minimum of one sample per leaf, two samples per split and optimal number of 158 estimators.

	In the next stage, for each classified spectrum within the test set, we computed individual probabilities derived from the RF ensemble. Specifically, for each classified spectrum, the model provides the probability of belonging to each of the two classes. In our binary model, classification is determined by a probability threshold of 0.5. Probabilities closer to unity indicate higher confidence that the pixel represents water spectra in various bloom intensities, while probabilities closer to zero indicate lower confidence that the pixel represents water or, alternatively, higher confidence that the pixel spectrum belongs to the non-water class. This confidence score is highly valuable for the subsequent Chl\textit{a} retrieval algorithm, as demonstrated below, where we show that the variance in Chl\textit{a} retrieval decreases with increasing classification confidence.

	\subsection{Dynamic Training Approach
	}
	
	\begin{figure*}[t]
		\includegraphics[width=\textwidth]{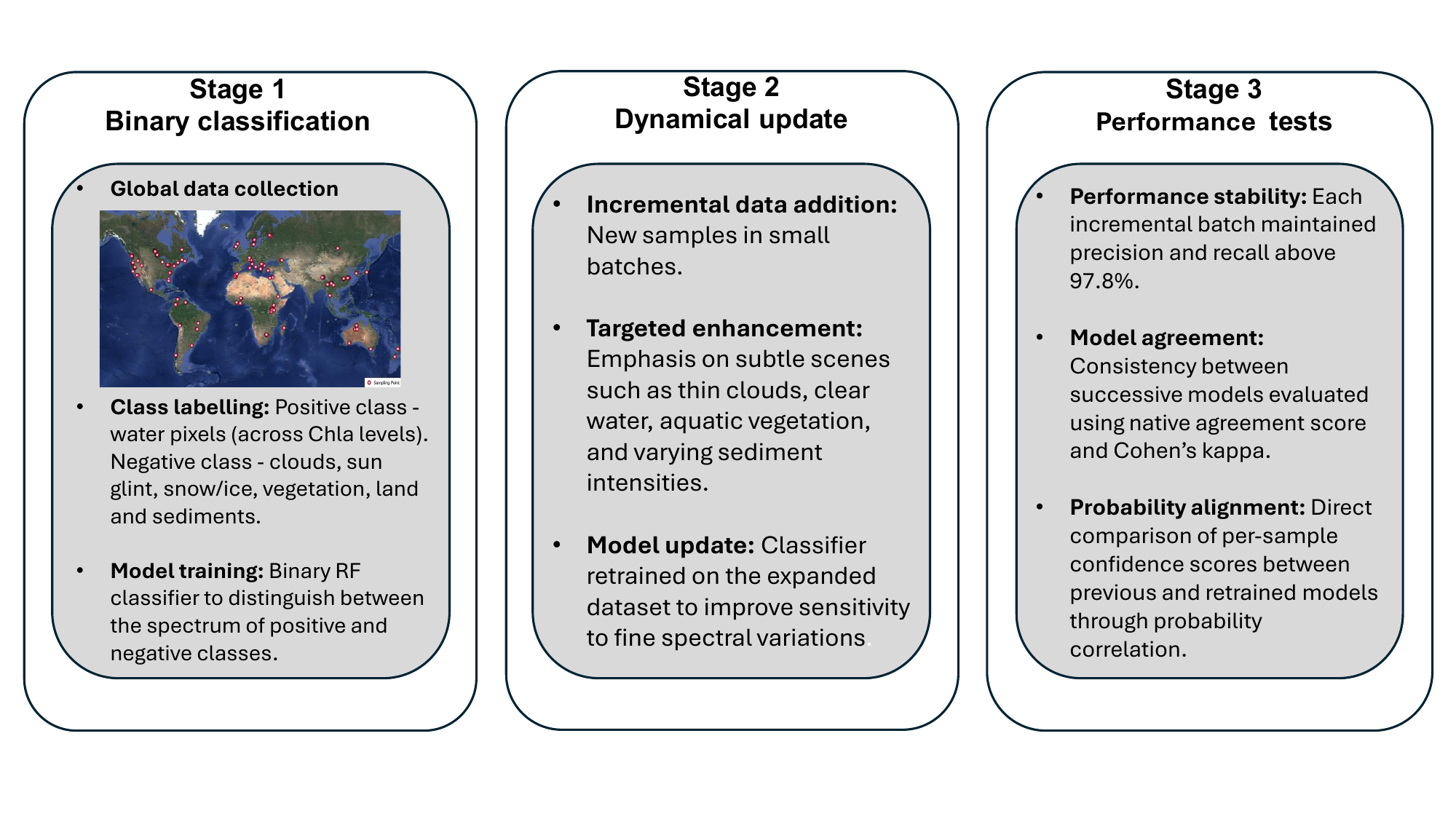}\vspace{-1.2cm}
		\caption{\small{Classification pipeline workflow. The scheme illustrates the iterative approach conducted throughout the training and updating process of the GWC.}}\label{gwcwf1}
	\end{figure*}
	The RF classifier employed a dynamic training approach, where new datasets were incrementally added in smaller batches. Each new dataset consisted of pixels from new water bodies and varying spectral and seasonal conditions, increasing the dataset's spatial and temporal diversity. While the initial training dataset was composed of well distinguished optical scenes like clouds, glint, ice and land, the new dataset mainly consists of subtle optical interference such as thin clouds, low concentrations of sediments and aquatic vegetation, as well as clear water whose spectral signature exhibits moderate to small deviation due to disruption in the atmospheric correction (see also workflow described in Fig.~\ref{gwcwf1}). Each scene spectrum was verified to ensure data quality. 
		After integrating each batch, the classifier was retrained from scratch to ensure that the model adapted effectively to new spectral signatures.

	Rigorous validation procedures were executed after each incremental retraining to ensure the stability and improvement of the classifier. These included cross-correlation analysis of predictions from successive model iterations to monitor shifts in classification patterns and confirm model consistency. Additionally, overall accuracy and confusion matrices were calculated to track classification improvements. For each sample, a probability score was calculated from the traditional trees aggregation. Then, a direct comparison of individual probability scores between previous and newly retrained models were conducted using Cohen’s kappa coefficient \cite{Cohen1960_kappa}, providing quantitative measurements of model agreement and reliability. Specifically, during retraining when new data was added, we closely monitored three key metrics: agreement rate, Cohen's kappa and probability correlation. 
	The agreement between two binary classifiers is defined as
	
	\begin{equation}
		\text{Agreement} = \frac{\sum_{i=1}^{N} \mathbf{1}\{y_i^{\text{old}} = y_i^{\text{new}}\}}{N},
	\end{equation}
	
	where $y_i^{\text{old}}$ and $y_i^{\text{new}}$ are the labels assigned by the old and new classifiers, respectively, and $\mathbf{1}\{\cdot\}$ denotes the indicator function.  
	
	Cohen’s kappa provides a chance-corrected measure of agreement
	
	\begin{equation}
		\kappa = \frac{p_0 - p_e}{1 - p_e},
	\end{equation}
	
	where $p_0$ is the observed agreement (the fraction of identical label samples) and $p_e$ is the expected agreement by chance (the random guess agreement).  
	
		We further assesses the  Pearson coefficient which quantifies the similarity of the predicted confidence scores between sequential model iterations
	by looking at the correlation between two sets of predicted confidence scores. 
	\begin{equation}
		\begin{aligned}
			r &= \frac{\mathrm{cov}(p_1, p_2)}{\sigma_{p_1} \, \sigma_{p_2}}\\
			&= \frac{\frac{1}{n}\sum_{i=1}^n (p_{1,i} - \bar{p}_1)(p_{2,i} - \bar{p}_2)}
			{\sqrt{\frac{1}{n}\sum_{i=1}^n (p_{1,i} - \bar{p}_1)^2} \cdot 
				\sqrt{\frac{1}{n}\sum_{i=1}^n (p_{2,i} - \bar{p}_2)^2}} ,
		\end{aligned}\label{pc}
	\end{equation}
	Where $p_1, p_2$ labels the per sample probability  of the old and the new binary classifiers.
	Throughout the retraining process, we not only maintained precision and recall scores of 97.8$\%$ and 98.8$\%$ respectively on the test set but also consistently achieved an agreement rate between successive models of 100$\%$, Cohen’s kappa values above 0.95, and probability correlations exceeding 95$\%$. To facilitate continuous improvement, detailed records of classifier performance metrics were maintained for each iteration. This dynamic methodology preserves the desired accuracy performances and the sensitivity to environmental changes, which eventually enhances the reliability of classification outcomes. The GWC workflow is presented in Fig.~\ref{gwcwf1}
	Having established a reliable classifier that effectively masks various optical and non-water interferences, we are now able to analyze clean pixels, which, as we demonstrate below, significantly enhances the accuracy of Chl\textit{a} concentration estimation.
	For our Chl\textit{a} retrieval algorithm, we utilized the extensive USGS 'AquaMatch' \cite{brousil2024_aquamatch} dataset, comprising more than 1 million geo-referenced in situ Chl\textit{a} measurements collected from diverse inland water bodies, including rivers, dams, lakes, reservoirs, and ponds spanning from the mid-1970s through 2024.
	From this extensive dataset, we randomly selected a subset of 20,000 measurements. Our selection criteria focused on data points collected after June 2018, aligning with the full operational capability of the Sentinel-2 satellite. Additionally, we limited our selection to water bodies classified specifically as 'Lake, Reservoir, or Impoundment' where water samples were taken at shallow depth of up to 1.5 meters.
	To ensure accurate alignment between satellite reflectance data and in-situ Chl\textit{a} measurements, we sampled each point using GEE within a 30-meter square grid. This spatial scale was specifically chosen to reduce potential GPS inaccuracies, minimize complications arising from points overlapping multiple pixels, and to agree with the data extraction from the relevant Sentinel-2 bands (B1–B12), which have spatial resolutions of 10–20 meters. For each 30-meter grid sample, we extracted the relevant Sentinel-2 bands and calculated the following indices: NDWI, NDVI, BI, FAI, and NDTI, in addition to the following spectral 
	\begin{gather}
		\text{CDOM}_{1} = \frac{R_{443}}{R_{560}}, \nonumber\\
		\text{CDOM}_{2} = \frac{R_{443}}{R_{490}}, \nonumber\\
		\text{Sediment}_{2} = \frac{R_{560}}{R_{665}}, \nonumber\\
		\text{Turbidity Index} = \frac{R_{560}}{R_{490}}, \label{ind1}\\
		\text{NDCI} = \frac{BI - 1}{BI + 1}, \nonumber\\
		\text{Three-band Chl\textit{a} Index type 1} = R_{740}\left(\frac{1}{R_{705}} - \frac{1}{R_{665}}\right),\nonumber\\
		\text{Three-band Chl\textit{a} Index type 2} = R_{783}\left(\frac{1}{R_{705}} - \frac{1}{R_{665}}\right)~.\nonumber
	\end{gather}
	Recognizing the Sentinel-2 temporal sampling characteristics, we extended the temporal matchup window to include satellite imagery acquired one day before or after each in-situ measurement date. This strategy significantly expanded the dataset size.
	Then we applied the GWC and filtered out non-water pixels, so only the water-labelled pixels were considered in the analysis. Where, again for clarity, by ‘water-labelled’ we mean to water pixels containing algal bloom imprint in various intensities. We average over their spectrum and created a new dataset containing in-situ Chl\textit{a} values along their corresponding spectrum.


	\section{The Three-stage retrieval algorithm}
	
	The Chl\textit{a} retrieval algorithm involves several key steps. The first is a filtering stage, where only valid and analyzable water pixels are retained through the GWC. The subsequent stages are designed to capture the non-linear relationship between water spectra and in-situ measured Chl\textit{a} concentrations. In ideal conditions, one might expect a linear or near-linear relationship, reflecting the inherent photic response of a single Chl\textit{a} pigment. However, in practice, this linearity breaks down due to multiple sources of interference such as natural factors (various water constituents), systematic effects (atmospheric correction models), and statistical noise (sensor uncertainty). These combined influences lead to inconsistencies in the observed optical properties across varying Chl\textit{a} concentrations, thereby highlighting the need for a robust non-linear modeling approach to capture these complex relationships. 
	\subsection{Stage 2 - Base Chl\textit{a} regression with XGBoost} \label{step2}
	We first train an XGBoost regressor to predict Chl\textit{a} from the input spectra and derived indices, using only pixels classified as valid water (“positive” class) we have left with  13,626 valid water pixels. 
	We split the dataset using a two-stage strategy that preserves spatial independence. First, a 70/30 train–test split is performed at the lake level, ensuring that each water body appears exclusively in either the training or testing set, thereby preventing spatial leakage and enabling a realistic assessment of generalization to unseen lakes. Within the training subset, we further apply grouped 5-fold cross-validation based on lake identity. In this scheme, lakes are divided into 5 folds, and the model is trained iteratively on 4 folds while validated on the remaining fold, with no lake shared across folds. This combined approach reduces overfitting, improves robustness, and ensures stable performance across geographically distinct water bodies.
The feature set is comprised of the above-mentioned spectral indices and bands. Additionally, for internal use and to ensure spatial heterogeneity we included the sampling coordinates and geolocation (lat/lon). 
	Model accuracy was evaluated on the held-out 30$\%$ test split using the statistical metrics listed below. Overall performance  reflects the model’s ability to capture the nonlinear spectral interaction and to recover the original Chl\textit{a} value Section.\ref{results}.  Nevertheless, residual analysis reveals systematic, regime-dependent increasing errors, for example at high Chl\textit{a} (captured by the NDVI and FAI) or under elevated CDOM/sediments (captured by the CDOM and sediment indices). The underperformance of the regression prediction highlights the need of an additional residual correction step, which aims to capture these conditions and reduce the prediction error.

	\begin{figure*}[t!]
		\includegraphics[width=\textwidth]{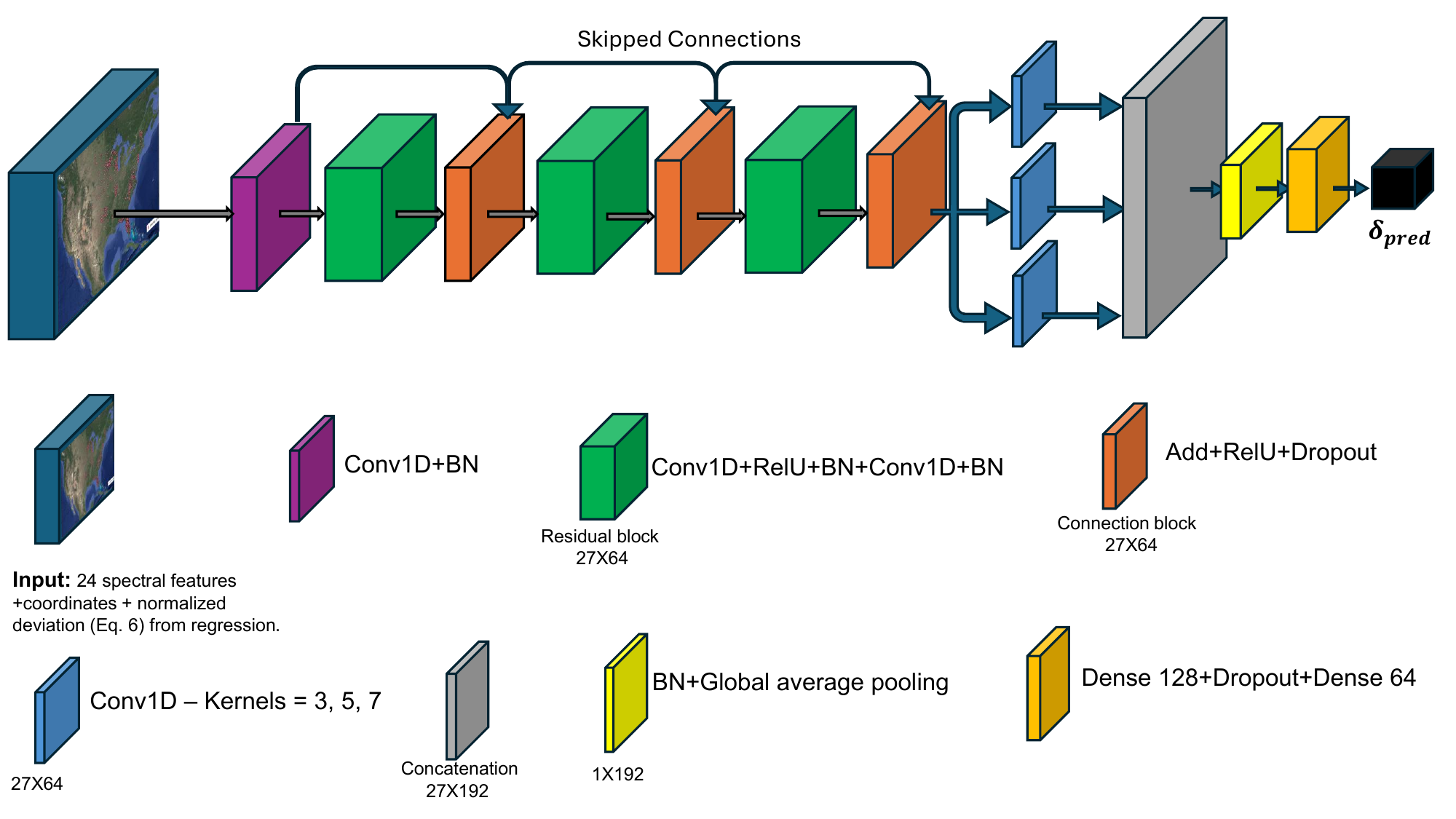}
		\caption{\small{Workflow of the residual correction architecture described in Sections. \ref{step3} and \ref{secres} } }
		\label{resnet}
	\end{figure*}
	\subsection{Stage 3 - Residual CNN correction model} \label{step3}
	After fitting the base XGBoost model, we construct a development set from the held-out test split by appending the model prediction as a feature ($y_{\text{pred}}^{\text{XGB}}$). The target feature is defined as the normalized deviation of the predicted Chl\textit{a} value with respect to the true in-situ sampled value. Specifically, the normalized deviation is
	
	\begin{equation}
		\delta = \frac{y_{\text{pred}}^{\text{XGB}} - y_{\text{true}}}{y_{\text{pred}}^{\text{XGB}}},
	\end{equation}
	
	where 
	\begin{equation}
		\Delta = y_{\text{pred}}^{\text{XGB}} - y_{\text{true}}
	\end{equation}
	
	is the raw prediction deviation, and $\varepsilon \ll 1$ is a small constant to prevent division by zero. Positive values of $\delta$ denote over-prediction and negative values under-prediction relative to the initial model. The development set is randomly divided into training and test partitions (50\%–50\%). All 27 input features are z-score standardized so that the 24-dimensional feature vector can be processed as a one-dimensional sequence.
	
	We then train a one-dimensional residual convolutional neural network (RCNN) to predict the target deviation $\delta$. The architecture is ResNet-like \cite{He2016_ResNet}, consisting of stacked Conv1D blocks with residual (skip) connections, branching into parallel Conv1D paths with varying kernel sizes to capture multi-scale spectral dependencies. The stacked residual blocks with skip connections aim to study gradually and efficiently the complex spectral patterns and to yield the corrected prediction.
	The logic is that in cases that the RCNN mapping is close to the identity, the residual block learn near zero weights. This scenario is suited to correction models where only a subtle adjustment to the original predictions are needed.
	Applying the residual CNN to the test set yields $\delta_{\text{pred}}$, from which the corrected Chl\textit{a} predictions are recovered as
	
	\begin{equation}
		\hat{y}_{\text{corr}} = y_{\text{pred}}^{\text{XGB}} \, (1 - \delta_{\text{pred}}).
	\end{equation}
	
	The RCNN was designed to capture spectral correlations at different regimes by varying the kernel size. More importantly, it was trained on residuals to learn the errors introduced by diverse interferences, primarily those governed by the water medium, and in some cases to mitigate artifacts from the Sen2Cor atmospheric correction, which is known to perform only moderately at low Chl\textit{a} concentrations \cite{WARREN2019267,rs15225370,rs14051099}. However, as shown below, compensating for Sen2Cor limitations is a more involved task that requires tailored, application-specific treatment, which we discuss in the following sections.
	
	The RCNN architecture is shown in Fig.~\ref{resnet}. In addition to the outlined residual structure, model parameters were optimized during training using the Adaptive Moment Estimation (ADAM) optimizer, with the Mean Squared Error (MSE) as the loss function.
	
	\subsection{Statistical evaluation metrics}
	
	We report accuracy in both linear and log spaces. Linear space preserves physical units and is intuitively interpreted, evaluated by R², MAE, and RMSE, and report the calibration slope from regressing the predicted Chl\textit{a} values  with respect to the truth values . To capture heteroscedastic (scale-dependent) behaviour and multirange performance, we also use log space and report RMSLE (and the corresponding slopes at both linear and log scales).
	
	\begin{gather}
		R^2 = 1 - \frac{\sum_{i=1}^{n} \left(y_i^{\text{true}} - y_i^{\text{pred}}\right)^2}
		{\sum_{i=1}^{n} \left(y_i^{\text{true}} - \overline{y^{\text{true}}}\right)^2} \\[6pt]
		\mathrm{MAE} = \frac{1}{n} \sum_{i=1}^{n} \left| y_i^{\text{true}} - y_i^{\text{pred}} \right| \\[6pt]
		\mathrm{RMSE} = \sqrt{\frac{1}{n} \sum_{i=1}^{n} \left( y_i^{\text{true}} - y_i^{\text{pred}} \right)^2} \\[6pt]
		\mathrm{RMSLE} = \sqrt{\frac{1}{n} \sum_{i=1}^{n} 
			\left( \log\!\left(y_i^{\text{pred}} + \varepsilon\right) - 
			\log\!\left(y_i^{\text{true}} + \varepsilon\right) \right)^2 }
	\end{gather}
	
	where $y_i^{\text{true}}$ and $y_i^{\text{pred}}$ denote the true and predicted 
	Chl\textit{a} concentrations for the $i$-th sample, $\overline{y^{\text{true}}}$ is the mean 
	of the true values, $n$ is the total number of samples, and $\varepsilon$ is a small 
	constant added to avoid undefined values when applying the logarithm in cases where prediction value is zero.

	

	\section{Results} \label{results}

	In the following, we present the results following the three-stage workflow, highlighting the algorithm’s performance and its behavior across varying scenarios.  In the second stage, the XGBoost regressor was trained on the dataset and achieved its best performance with hyperparameters of 200 estimators and 0.1 learning rate. The corresponding accuracy metrics and scatter plots are presented below. A key advantage of the GWC approach lies in its ability to distinguish and exclude non-water pixels that are affected by strong optical interferences. The degree of clarity is quantified through the probability measure output by the RF classifier, which provides a pixel-based estimation of water detectability. The probability measure of valid water pixels serves as an additional accuracy measure that reduces the variance of the predicted Chl\textit{a} compared to the true values. Pixels covered by clouds, land, ice, and snow generally receive very low probability score (near zero) whereas other interferences, such as strong glint, pronounced sediments and aquatic vegetation, gets higher probability score but are still labeled as no.
	
	\begin{figure}[t!]
		\centering
		\includegraphics[width=1\linewidth]{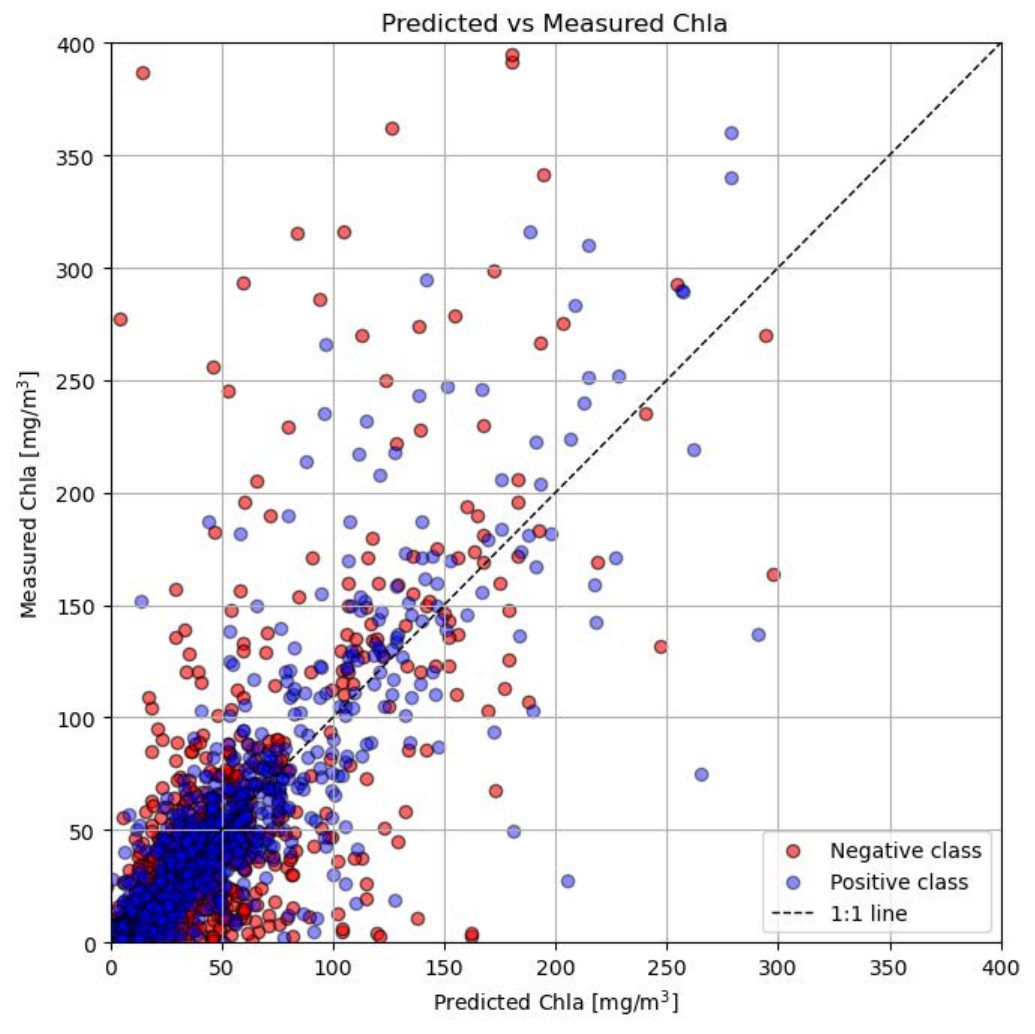}
		\caption{\small{Chl\textit{a} retrieval performances of the base XGB regression model.
				A subsample of 1114 datapoints from each class is shown. 
				The performances of the positive class significantly outperform the negative one, 
				as summarized in Table~\ref{tab:metrics_class}.}}\label{scatter_xgb}
	\end{figure}
	
	\begin{figure*}[t!]
		\includegraphics[width=\textwidth]{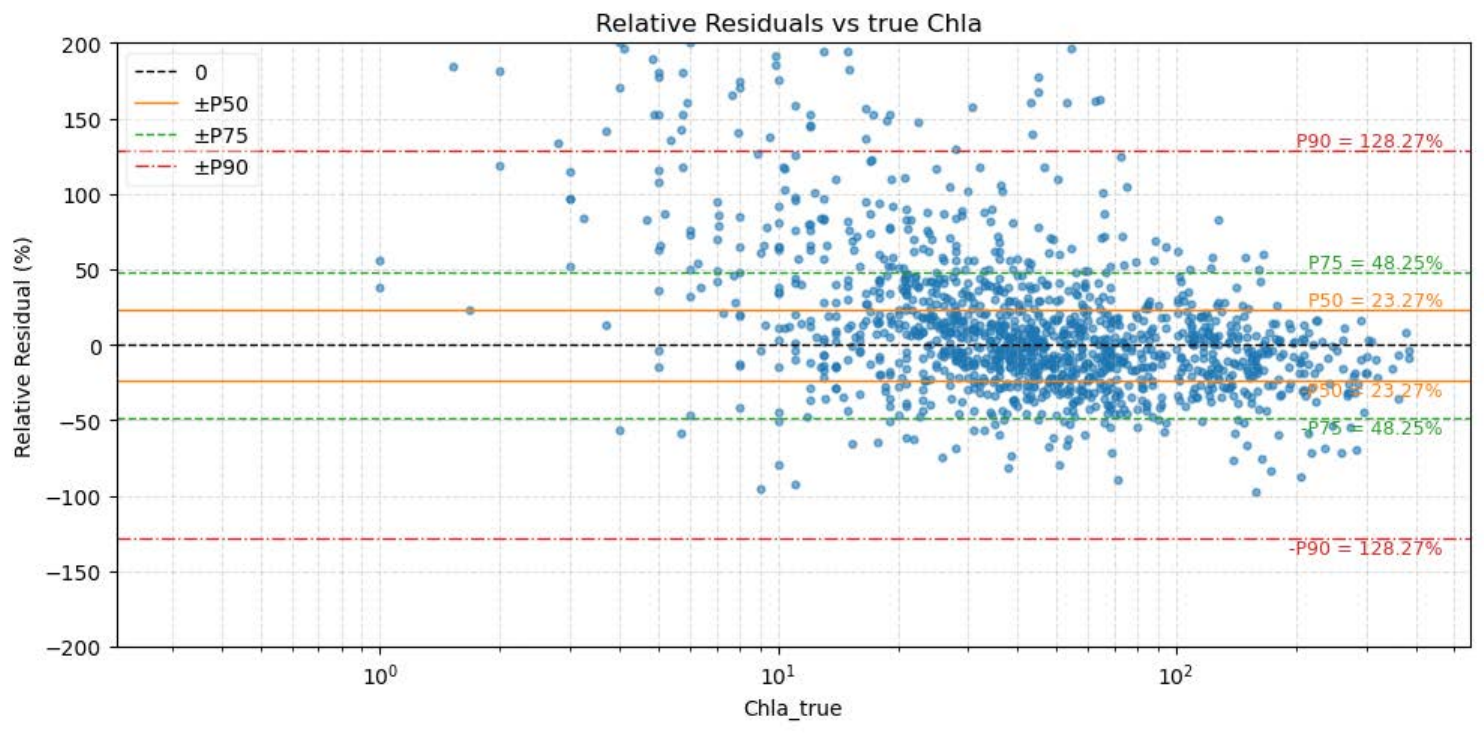}
		\caption{\small{The relative residuals are calculated by $ (y^{\text{pred}}_{\text{XGB}} - y^{\text{true}})/y^{\text{true}}$}. The 50, 75 and 90 percentiles are shown. The relative residuals increase as Chl\textit{a} values decrease. Quantitative assessment of the spread behavior is summarized in Table. \ref{tab:cv_entropym}.} 
		\label{relres1}
	\end{figure*}
	
	To illustrate the improved performance of the XGBRegressor when applied to the classified data through the GWC, we present a scatter plot of the two classes - predicted vs true Chl\textit{a} values for the positive and negative classes (near zero probabilities were excluded) and calculate accuracy metrics with respect to the true value. We generated a balanced dataset by considering an identical amount of ‘negative’ and 'positive' class samples Fig. \ref{scatter_xgb}.

	\begin{table}[t]
		\centering
		\small
		\setlength{\tabcolsep}{2.5pt}
		\renewcommand{\arraystretch}{1.15}
		\begin{tabular}{lccc}
			\toprule
			& \textbf{Positive class} & \textbf{Negative class} & \textbf{Rel.\ Dev.\ (\%)} \\
			\midrule
			${R}^\mathbf{2}$ & 0.63 & 0.37&70.27 \\
			MAE   & 12.62  & 31.14&59.47 \\
			RMSE  & 34.03  & 65.30&47.89 \\
			RMSLE & 0.585  & 0.944&38.03 \\
			MAPE  & 110.91\% & 198.16\%&44.03 \\
			\bottomrule
		\end{tabular}
		\caption{Regression performance on the positive and negative classes.}
		\label{tab:metrics_class}
	\end{table}
	The results presented in Table.\ref{tab:metrics_class} show a significant improvement in retrieval performances of the positive over the negative class. The results emphasize the advantage  of the probability-based classification method, where $p=0.5$ serves as the threshold where points above and below are classified as positive and negative, respectively. 
	\subsection{Residual analysis}
	Residual analysis reveals that the deviations between predicted and observed values are not purely random but instead exhibit a structured pattern. Two distinct regimes can be identified, separated by a Chl\textit{a} concentration of approximately $\approx 10 mg/m^3$. At lower concentrations (Chl\textit{a} $< 10 mg /m^3$), relative residuals increase as concentrations decrease. This behavior is primarily attributed to limitations in the Sen2cor atmospheric correction, which is less effective in retrieving reflectance from optically clear waters. As a result, even small concentration differences are amplified, leading to prediction errors that manifest as random or noise-like variability, similar behavior was reported in \cite{rs15225370,rs14051099}. In contrast, at higher concentrations (Chl\textit{a}$ > 10 mg /m^3$), prediction errors display a more systematic structure rather than random scatter. This residual behavior is evident in Fig.\ref{relres1}.
	
	To demonstrate the residual structure, we use two metrics. The first and the more intuitive is the coefficient of variance (CV), which is the ratio of the standard deviation $\sigma$ to the mean $\mu$ 
	\begin{equation}
		CV = \dfrac{\sigma}{\mu}
	\end{equation}
	This dimensionless quantity measures the relative variability of data around the mean, and how scatter is the data around it. 
	
	\begin{figure*}[t!]
		\includegraphics[width=0.75\textwidth]{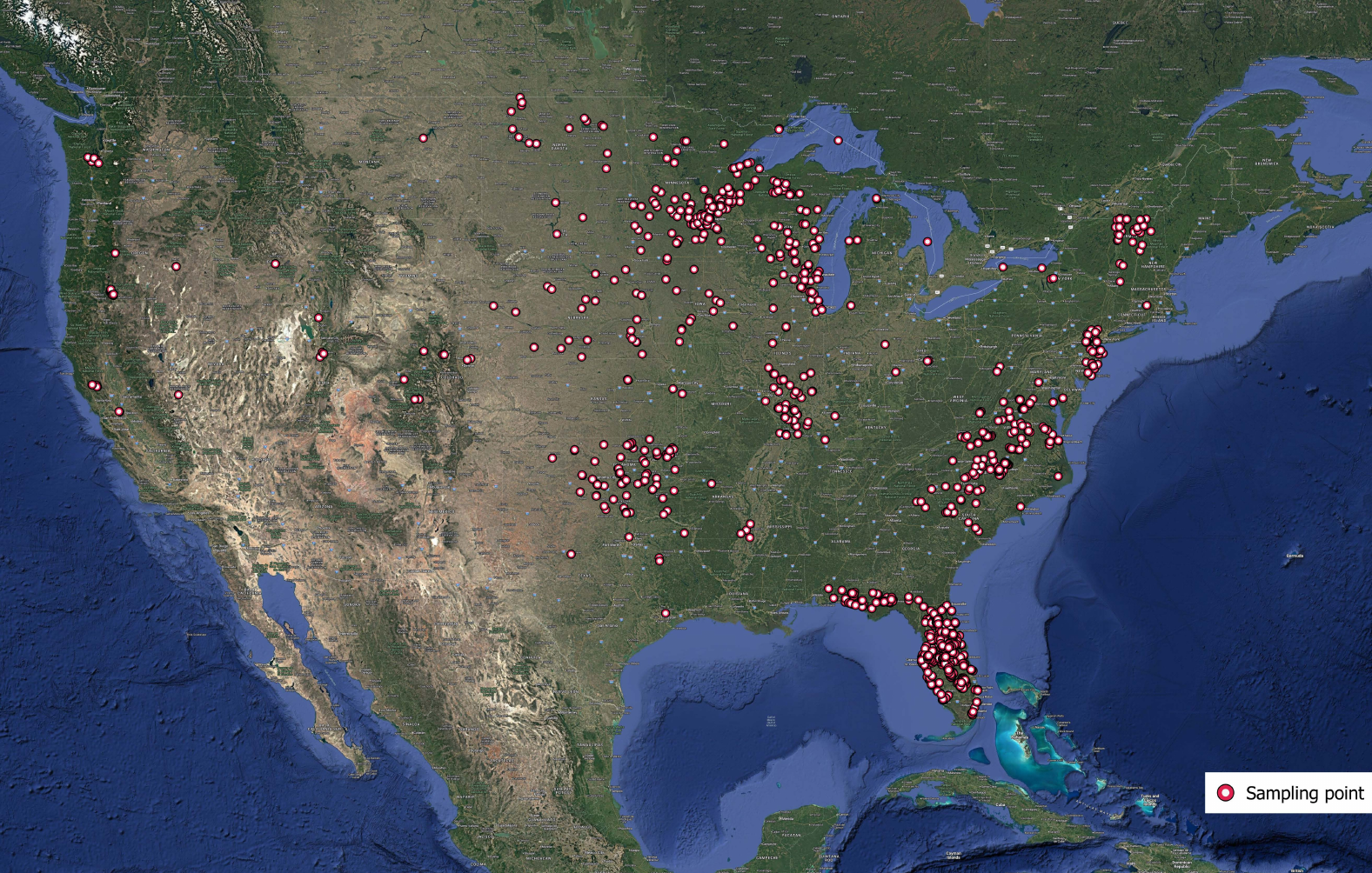}
		\caption{\small{Spatial spread of the residual CNN test set. Data extracted from the USGS ’AquaMathc’ dataset.
				Sample points where selected specifically as ’Lake, Reservoir, or Impoundment’ and depth below 1.5 meter. The test set covers 867 different water bodies.} }
		\label{world_test}
	\end{figure*}
	The second quantity is the Shannon entropy, which measures the degree of randomness or uniformity in the residual distribution. Unlike CV, which focuses on spread, entropy is sensitive to how evenly the residuals are distributed across their possible values. Higher entropy means the residuals are more uniformly distributed (less structure), while lower entropy suggests concentration in certain values or patterns.
	The calculation of the Shannon entropy is performed by standardizing the data and taking the z-scored residuals . Let $r_i = y_{pred}-y_{true}$ denote the residuals for sample $i$, where ${y}_{pred}$ and $y_{true}$ are the predicted and true values, respectively. To remove scale biases, residuals are z-scored
	\begin{equation}
		z_i = \frac{r_i - \mu_r}{\sigma_r},
	\end{equation}
	where $\mu_r$ and $\sigma_r$ denote the mean and the standard deviation of the residuals. The standardized residuals were then converted into a probability distribution $p_j$ by binning into $k$ intervals, such that $p_j = \tfrac{\text{count in bin } j}{n}$ with $\sum_{j=1}^k p_j = 1$. The Shannon entropy is then given by
	\begin{equation}
		H = - \sum_{j=1}^{k} p_j \, \log(p_j),
	\end{equation}
	For comparability across different choices of $k$, the entropy is normalized to the interval $[0,1]$ by $H_{\text{norm}} = {H}/{\log(k)}$.
	In this way, the z-score normalized entropy provides a scale-invariant measure of the spread of the residual distribution.
	When combined, the coefficient of variation (CV) and entropy provide a more complete description of the residual structure, as CV quantifies the magnitude of variability and the entropy evaluates the degree of randomness in that variability.

	\begin{table}[ht]
		\centering
		\renewcommand{\arraystretch}{1.2} 
		\setlength{\tabcolsep}{12pt}      
		\begin{tabular}{lcc} 
			\toprule
			& \textbf{CV} & \textbf{Entropy} \\
			\midrule
			Chl\textit{a} $< 10$   & 3.07 & 0.73 \\
			Chl\textit{a} $> 10$   & 0.87 & 0.36 \\ 
			\bottomrule
		\end{tabular}
		\caption{\small{Coefficient of variation (CV) and entropy values computed for residuals above and below 10 mg /m$^{3}$.}}
		\label{tab:cv_entropym}
	\end{table}

	The results in Table \ref{tab:cv_entropym} show a moderate systematic error for Chl\textit{a} concentrations above 10 mg$/$m$^{3}$, and a more noise-like error pattern at lower values. Notably, entropy is not low or near zero in either regime, consistent with noise-like behavior across the dataset. This indicates that the residual structure may be better captured by a more sophisticated method, which will be discussed in the following section.

	\subsection{Residual correction algorithm}\label{secres}
	As described in Section~\ref{step3}, the residual correction model is applied via a CNN and aim to refine the predictions of the base regression model. The input to this model consisted of the same spectral and index features used in Step~\ref{step2}, together with the XGBoost prediction feature $y^{\text{pred}}_{\text{XGB}}$. The RCNN thus learned to model the deviations of the XGBoost predictions from the in-situ values rather than the Chl\textit{a} concentrations directly.

	\begin{figure}[t!]
		\centering
		\includegraphics[width=1\linewidth]{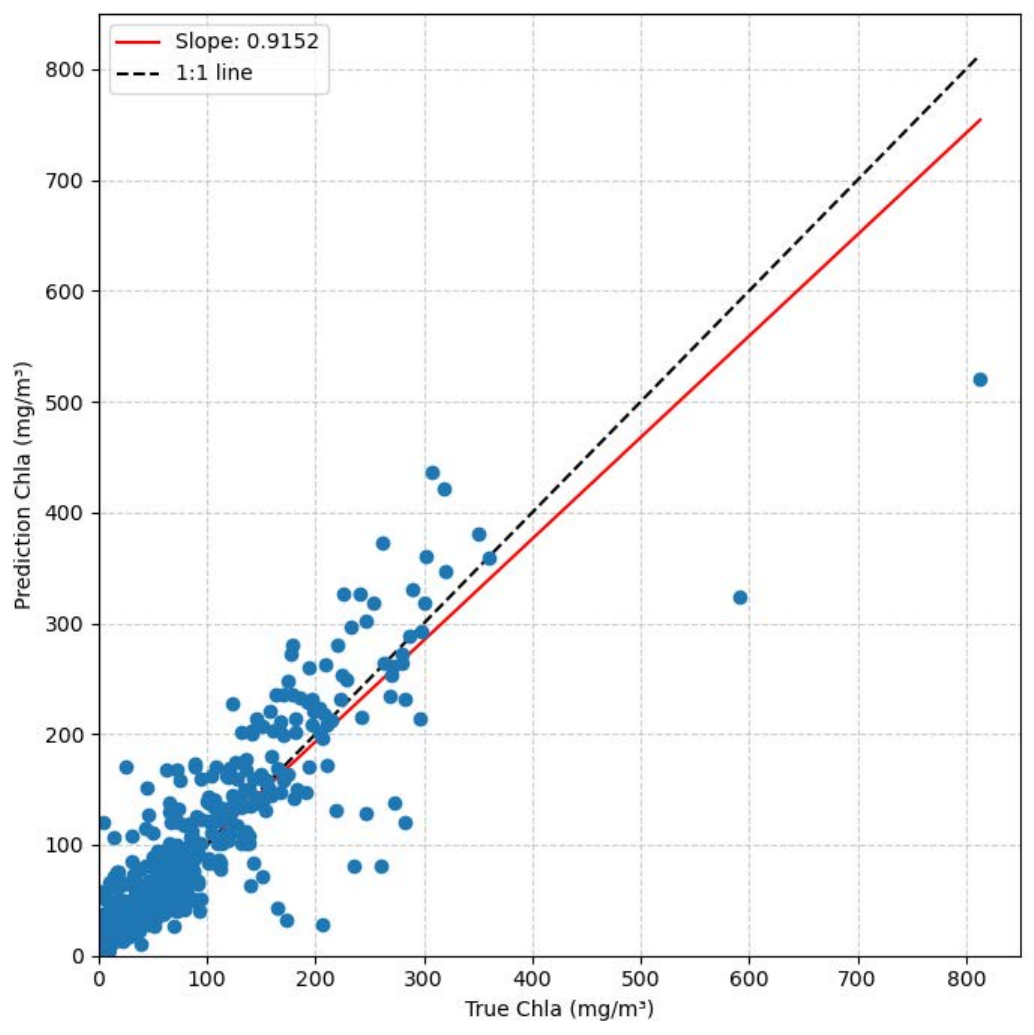}
		\caption{\small{Linear-scaled prediction plot based on 2,044 sampling points. The 1:1 line represents the ideal Chl\textit{a} retrieval, while the red solid line denotes the fitted linear regression.}}\label{linear_all}
	\end{figure}
	
	As shown in the previous section, the prediction of low Chl\textit{a} values ($\lesssim 10 \,\text{mg}/\text{m}^{3}$) is noisier and carries higher uncertainty
	 because the error structure differs significantly above and below this threshold, we train the correction algorithm on each segment independently. Accordingly, the regression predictions are divided into two groups: below and above $10 \,\text{mg}/\text{m}^{3}$, then the RCNN is applied on each regression output separately. 	

	To ensure independence between training and evaluation, the regression train-test sets were divided into two equal parts\footnote{The original dataset was linearly balanced across all regimes (nearly same number of samples at each bin of 10mg$/ $m$^3$), we increased the dataset size in the low concentration regime to have a log-scaled balanced at multiples of 10.}.
	In the following, we report the performance of the CNN residual correction on its test set
	
\begin{figure}[b!]
		\centering
		\includegraphics[width=1\linewidth]{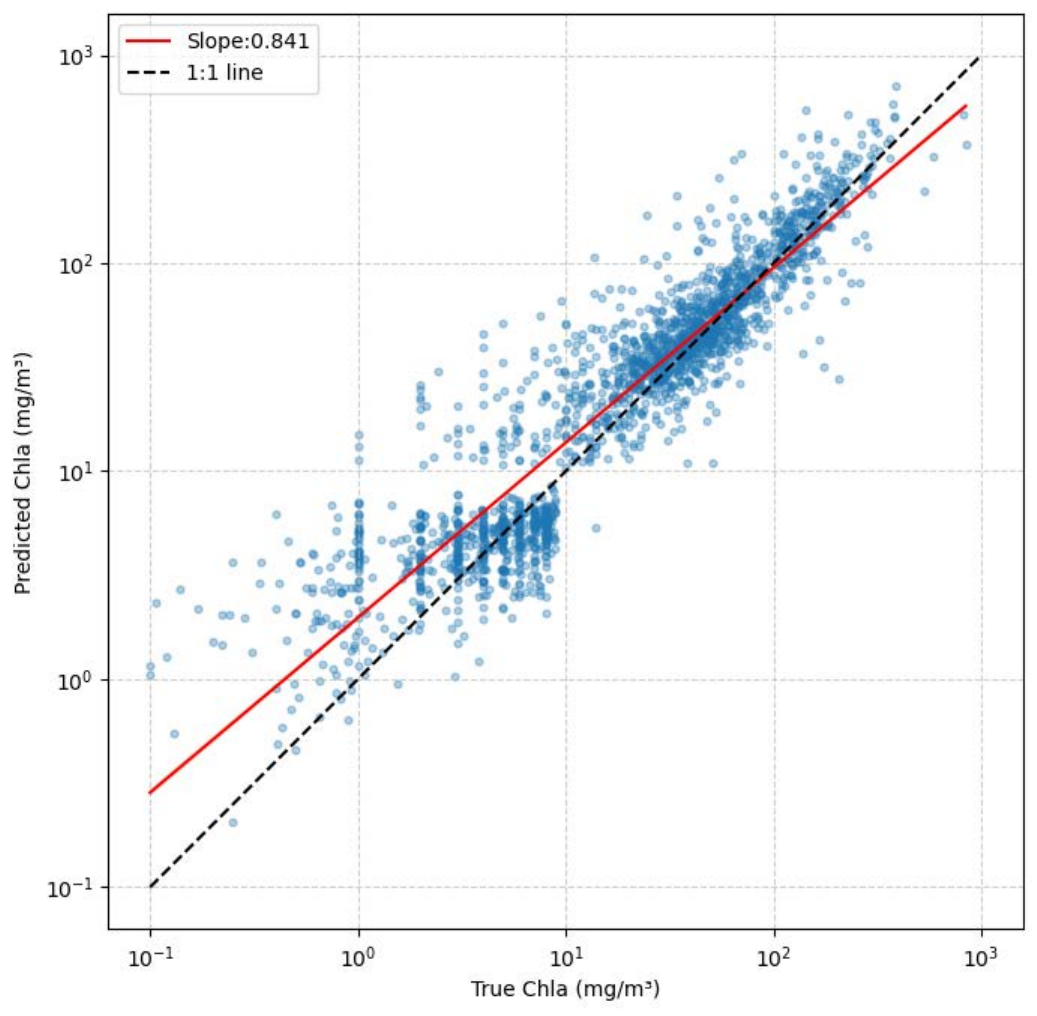}
		\caption{\small{Log-scaled prediction plot based on 2,044 sampling points. The 1:1 line represents the ideal Chl\textit{a} retrieval, while the red solid line denotes the fitted linear regression. The gap around Chl\textit{a} = 10 $mg/m^3$
				results from the post-regression splitting stage, in which the residual CNN algorithm is trained separately on predictions above and below this threshold. }}
		\label{log_all}
	\end{figure}

	We display both linear Fig.~\ref{linear_all} and log Fig.~\ref{log_all} scale results to capture relative and absolute deviations. The number of points analyzed $n=2044$, the geographical spread is displayed in Fig.\ref{world_test}.
	\begin{table}[h!]
		\centering
		\small
		\setlength{\tabcolsep}{2.5pt}
		\renewcommand{\arraystretch}{1.15}
		
		\label{tab:reg_perf}
		\begin{tabular}{lccc}
			\toprule
			& \textbf{Corrected} & \textbf{Base Regression} & \textbf{ Rel.\ Dev.\ (\%)} \\ 
			\cmidrule(lr){1-4}
			$R^{2}$ & 0.79 & 0.69 & 12.66 \\
			Slope   & 0.91 & 0.74 & 18.68 \\
			MAE     & 13.52 & 20.07 & 48.45 \\
			RMSE    & 26.63 & 35.89 & 34.77 \\
			RMSLE   & 0.56 & 0.60 & 7.14 \\
			\bottomrule
		\end{tabular} \caption{Regression performance before and after the CNN correction on the linear scaled data. Relative deviation is given is absolute values.}
		\label{relres3} 
	\end{table}
	
	The accuracy metric outlined in Table \ref{relres3} indicates a substantial improvements of the residual CNN algorithm compared to the base XGBoost regression, which suggest that overall, the correction step reduces the error magnitude.

	\begin{table}[h!]
		\centering
		\setlength{\tabcolsep}{12pt}      
		\begin{tabular}{lcc}
			\toprule
			& \textbf{CV} & \textbf{Entropy} \\
			\midrule
			Chl\textit{a} $< 10$   & 1.85 & 0.50 \\
			Chl\textit{a} $> 10$   & 0.85 & 0.34 \\ 
			\bottomrule
		\end{tabular}
		\caption{  Residual pattern after the application of the residual CNN correction algorithm. }
		\label{tab:cv_entropy_2}
	\end{table}
	The residual correction algorithm substantially reduces CV and entropy at low concentrations, with only minor improvements at higher levels as suggested in Table. \ref{tab:cv_entropy_2}. Indicating that the error correction algorithm captured the error structure, especially at low concentration levels and, as expected, justifies the application of the correction step. An examination of the metrics presented in Tables \ref{relres3} and \ref{tab:cv_entropy_2} demonstrates that the tailored residual treatment effectively reduces both the magnitude of the error and its variability, as well as the associated dispersion.
	
	The log-scaled results are presented in Fig.~\ref{log_all}, with the corresponding accuracy metrics summarized in Table~\ref{log_table}. Similar to the linear-scaled case, the performance improvements are most pronounced in the absolute metrics (slope and MAE), while the relative metrics exhibit only modest gains.
	
	\begin{table}[t!]
		\centering
		\small
		\setlength{\tabcolsep}{2.1pt}
		\renewcommand{\arraystretch}{1.15}
		\label{tab:reg_perf_case}
		\begin{tabular}{lccc}
			\toprule
			& \textbf{Corrected} & \textbf{Base Regression} & \textbf{ Rel.\ Dev.\ (\%)} \\
			\cmidrule(lr){1-4}
			$R^{2}$ & 0.75 & 0.68 & 9.33 \\
			Slope   & 0.84 & 0.67 & 20.24 \\
			MAE     & 15.67 & 18.56 & 18.44 \\
			RMSE    & 36.78 & 38.31 & 4.16 \\
			RMSLE   & 0.51 & 0.53 & 3.92 \\
			\bottomrule
		\end{tabular}
		\caption{\small{Prediction performances before and after the CNN correction on the log-scaled data. Relative deviation is given is absolute values.}}
		\label{log_table}
	\end{table}

	
	\begin{figure*}[!t]
	\centering
	\includegraphics[width=\textwidth]{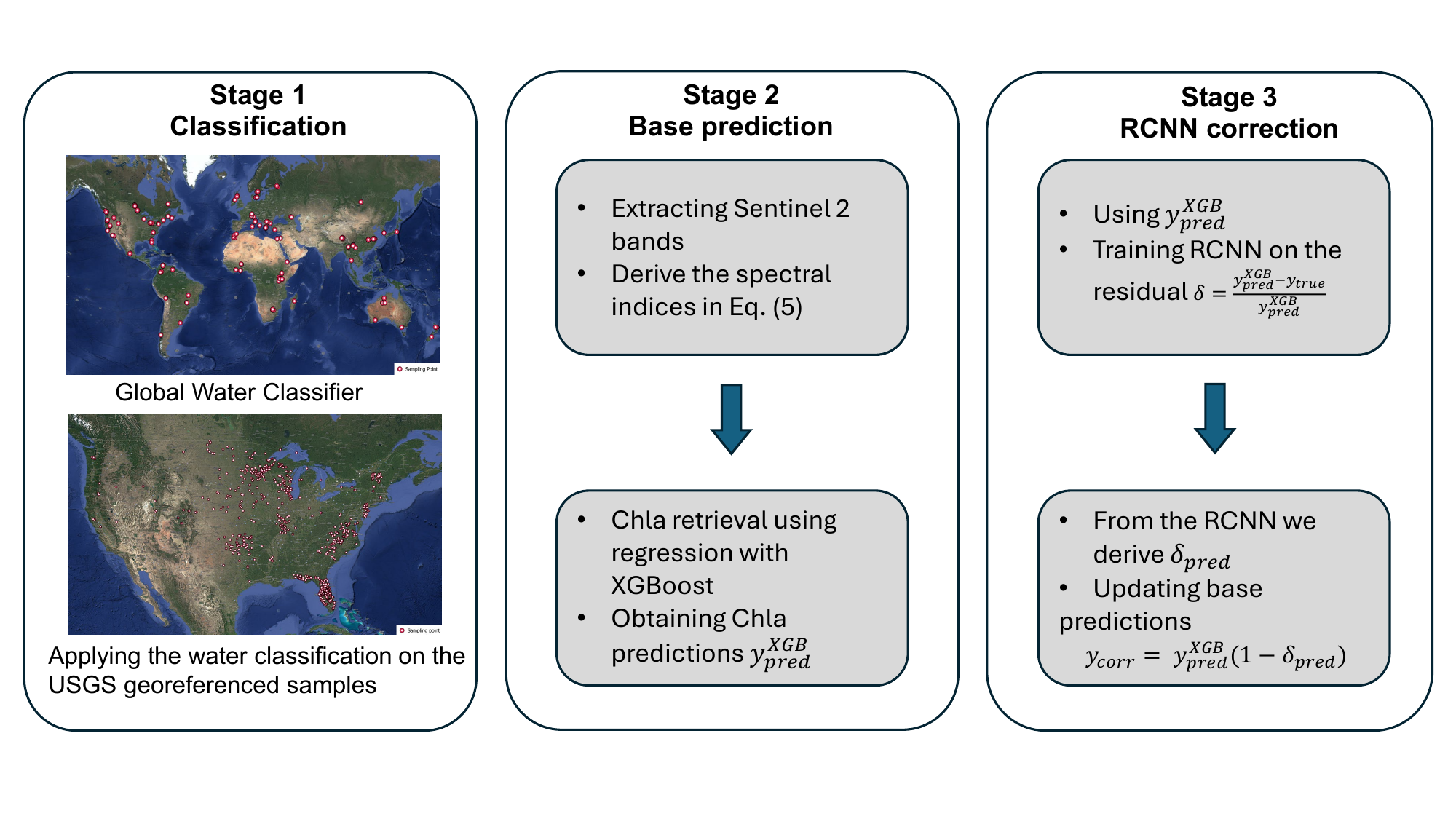}
	\caption{\small Workflow of the three-stage prediction algorithm.}
	\label{world_test1}
\end{figure*}

	\section{Full workflow – test and illustration}\label{WF1}

\begin{figure*}[!htbp]
		\centering
		\includegraphics[width=\textwidth]{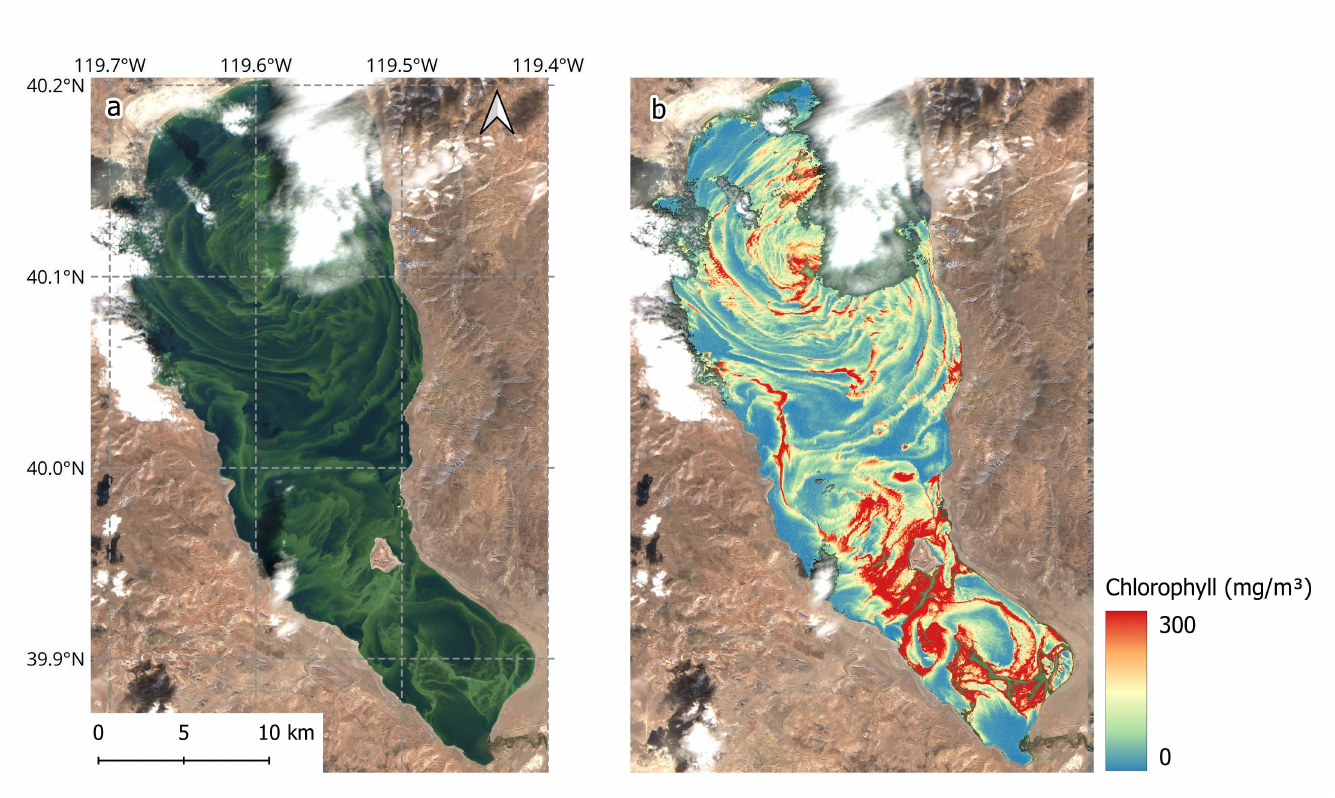}
		\caption{\small Pyramid Lake, Nevada, USA. a) RGB image, b) Chl\textit{a} prediction. Clouds removal by the GWC classifier. HAB at various densities is well captured.}
		\label{fig:pyramid}
	\end{figure*}

\begin{figure*}[!htbp]
		\centering
		\includegraphics[width=\textwidth]{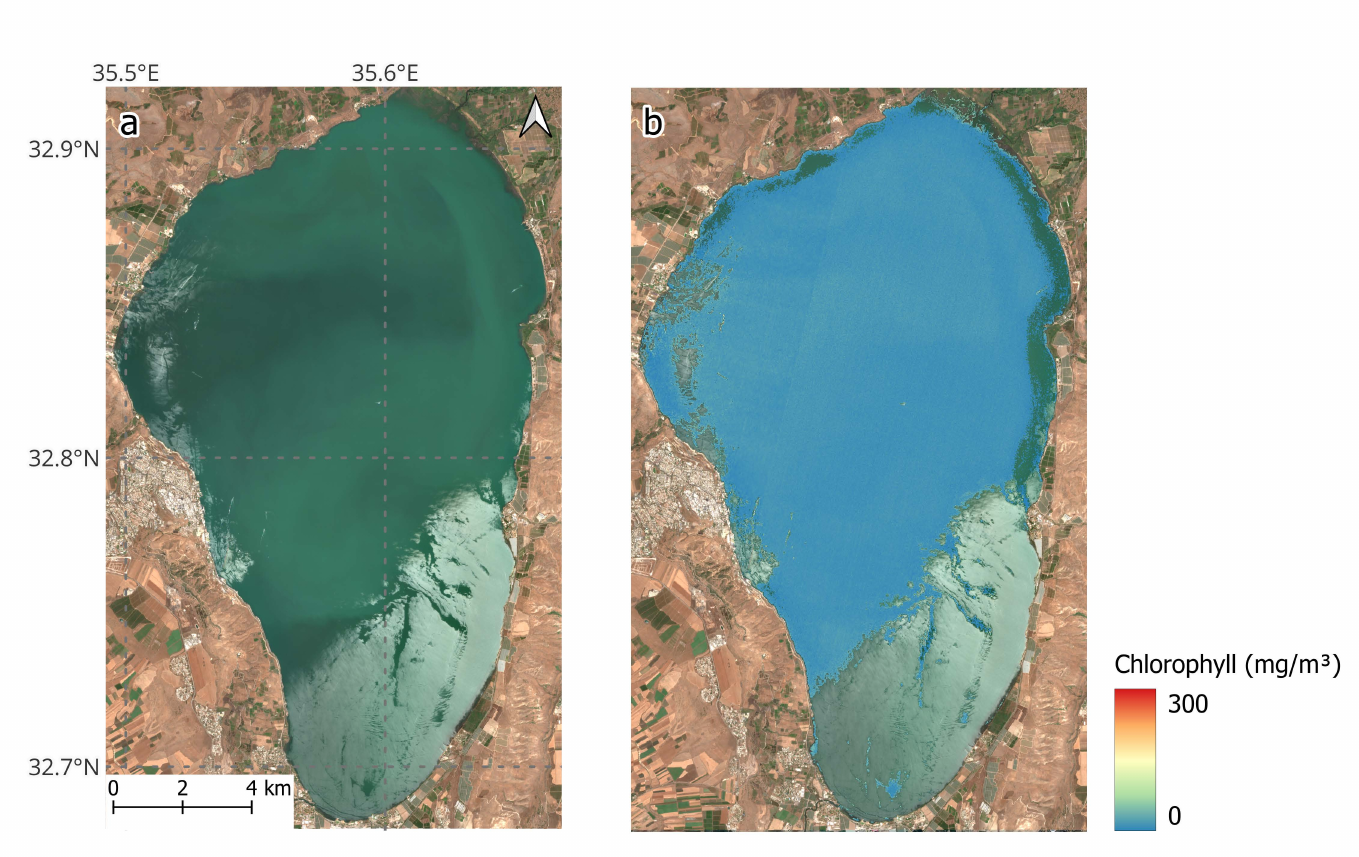}
		\caption{\small Kineret Lake, Galilee, Israel.  a) RGB image, b) Chl\textit{a} prediction. Sun-glint removal at various intensities.}
		\label{fig:kineret}\vspace{1.1cm}
	\end{figure*}

\begin{figure*}[!htbp]
		\centering
		\includegraphics[width=\textwidth]{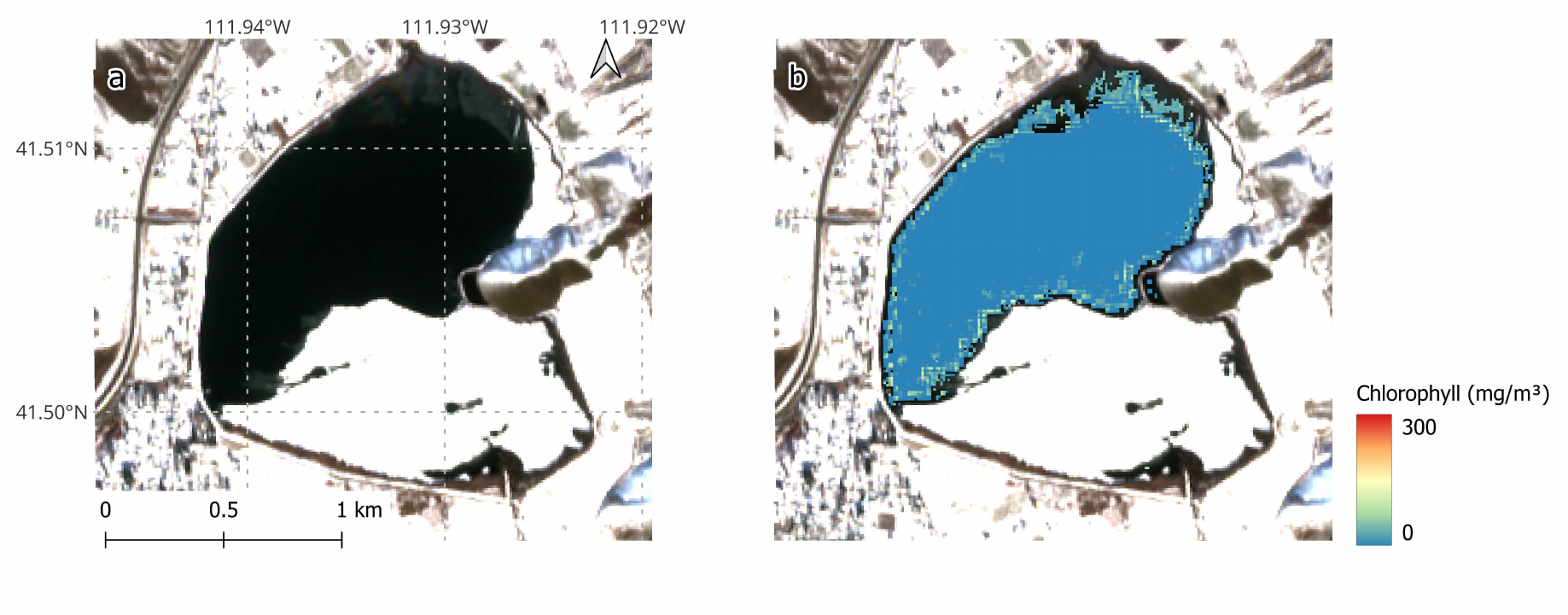}
		\caption{\small Mantua Reservoir, Utah, USA,   a) RGB image, b) Chl\textit{a} prediction. Ice, thin ice, and snow masking by the GWC.}
		\label{fig:mantua}
	\end{figure*}

\begin{figure*}[!htbp]
		\centering
		\includegraphics[width=\textwidth]{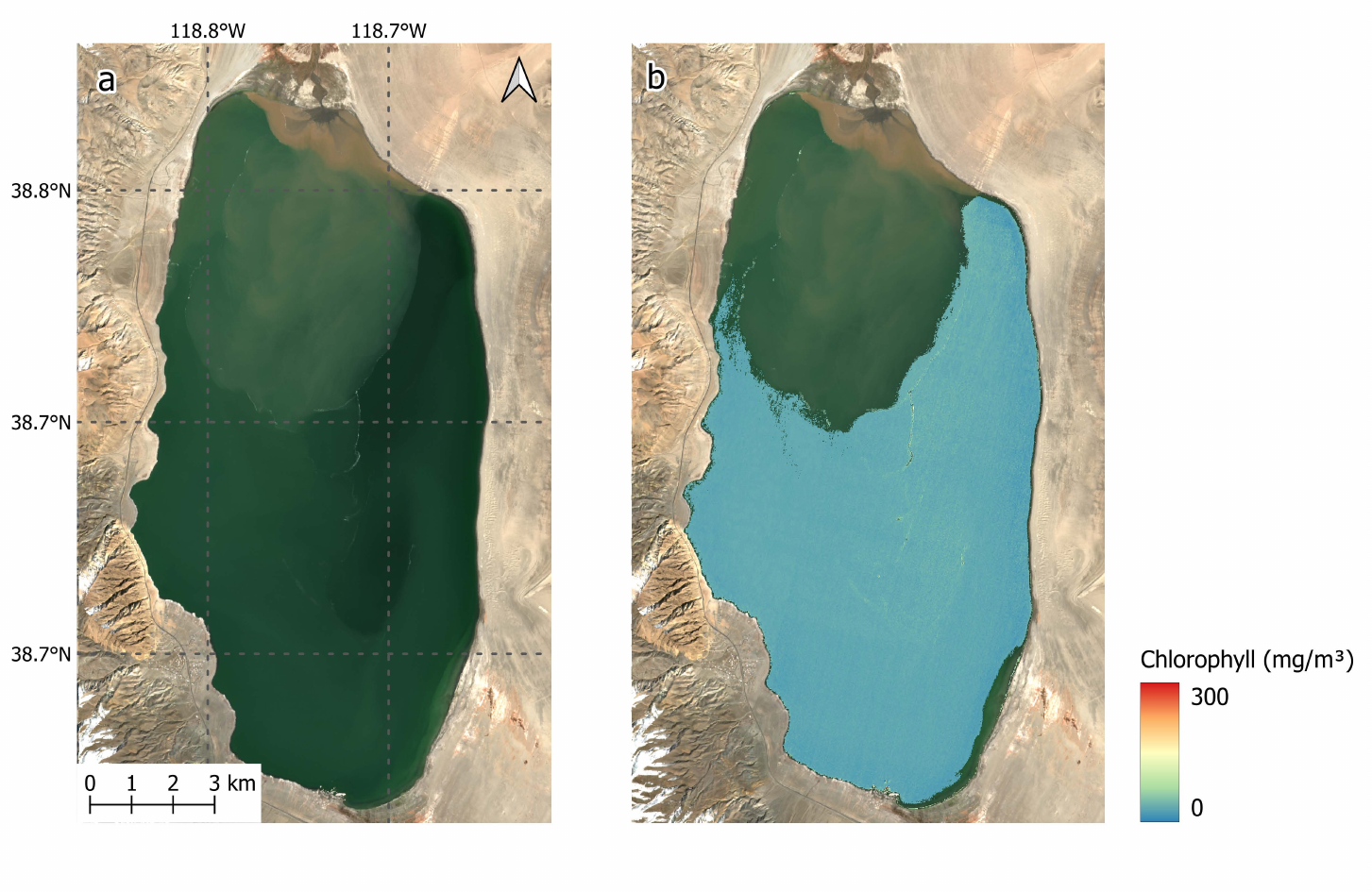}
		\caption{\small Walker Lake, Nevada, USA.   a) RGB image, b) Chl\textit{a} prediction. Sediment removal by the GWC classifier.}\vspace{1.1cm}
		\label{fig:walker}
	\end{figure*}

\begin{figure*}[!htbp]
		\centering
		\includegraphics[width=\textwidth]{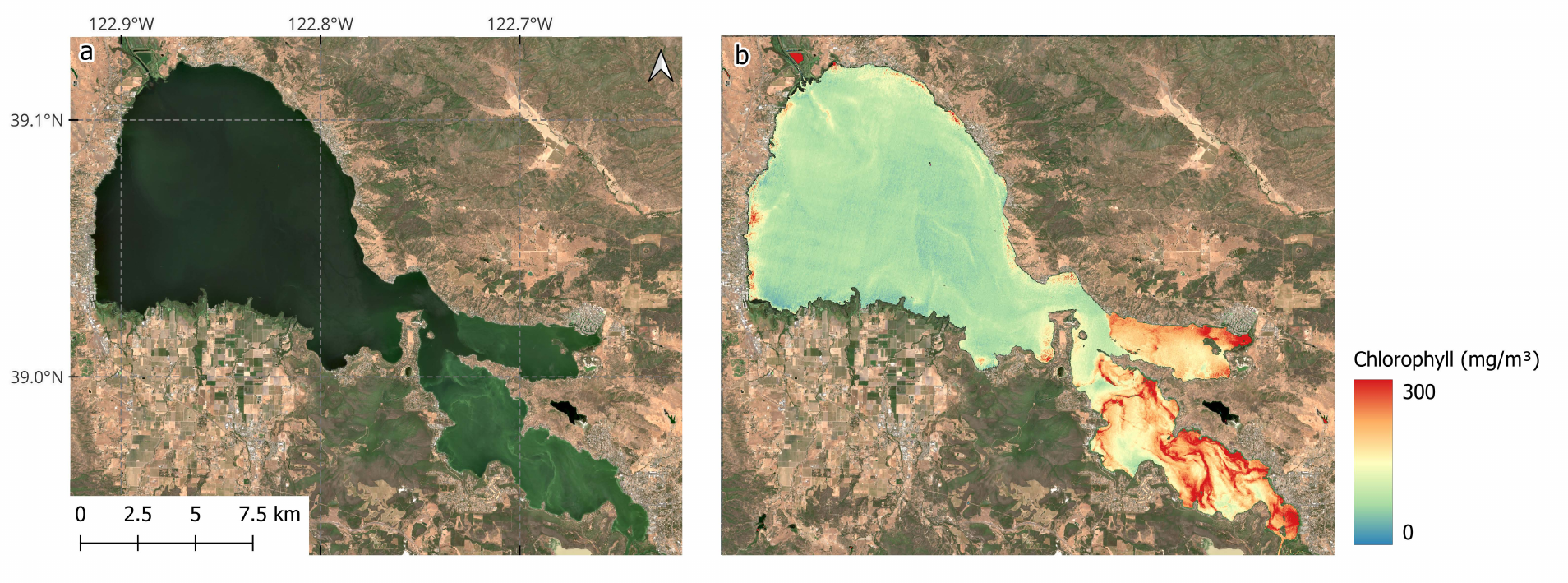}
		\caption{\small Clear Lake, California, USA.   a) RGB image, b) Chl\textit{a} prediction. HAB detected at various intensities.}
		\label{fig:clear}
	\end{figure*}

\begin{figure*}[!htbp]
		\centering
		\includegraphics[width=\textwidth]{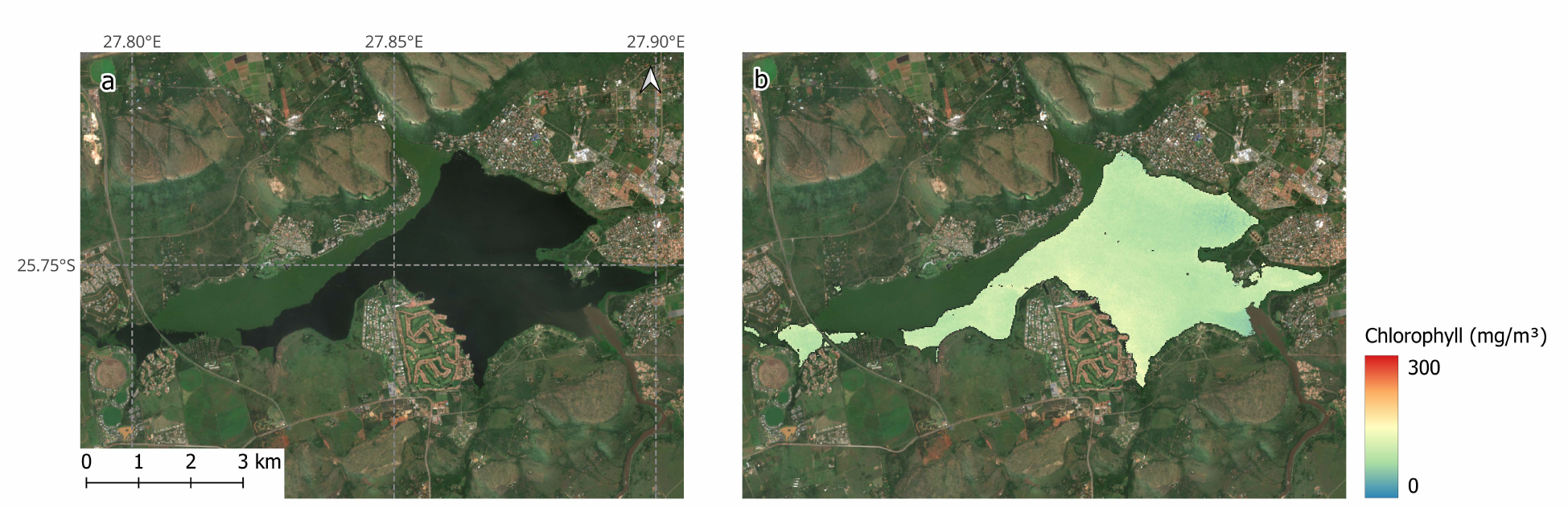}
		\caption{\small Hart Lake, Oregon, USA.   a) RGB image, b) Chl\textit{a} prediction. Water vegetation and sediment (bottom-right channel) removal.}
		\label{fig:hart}
	\end{figure*}

	For completeness of the process, we summarize the overall workflow. In Fig.~\ref{world_test1} we outline the three-stage Chl\textit{a} prediction algorithm, starting from the GWC and the subsequent regression and residual mitigation.
	Then, we demonstrate the full pipeline performances on several water bodies in different scenarios by considering several optical interference, mainly clouds  Fig.~\ref{fig:pyramid}, sun-glint Fig.~\ref{fig:kineret}, ice and snow Fig.~\ref{fig:mantua}, sediments Fig.~\ref{fig:walker}, aquatic vegetation Fig.~\ref{fig:hart} and varying Chl\textit{a} intensities Fig.~\ref{fig:clear}

	\section{Summary $\&$ Discussion}
	We developed a supervised Global Water Classifier (GWC) trained on Sentinel-2  surface reflectance data corrected via the Sen2Cor atmospheric process. We trained a binary RF classifier to distinguish between water at various Chl\textit{a} levels to non-water spectra (clouds, glint, snow, ice, vegetation, land and sediments). We use Sentinel-2 bands (B2–B12) and additional spectral indices Eq.\ref{ind1}, which facilitates the classification. The collected data comprises around 500 matched observations from nearly 100 lakes/reservoirs globally distributed (see distribution in Fig.\ref{globalmap}). The resulting classifier exhibits geographically stable performance, with uncertainties arising in optically complex scenes and especially in cases affected by atmospheric-correction artifacts. 
	In the next stage, we employed the GWC as a first stage in a 3-stages Chl\textit{a} retrieval algorithm. In the second stage we applied the GWC over the USGS AquaMatch in-situ dataset and extracted Sentinel-2 L2 surface reflectance for the selected sampling points. In addition to the required L2 bands and the derived indices Eq.\ref{si12}, we further computed the additional indices stated in Eq.\ref{ind1} to catch more subtle patterns. We then trained an XGBregressor on both the positive and negative classified pixels and witnessed the superiority of the positive class performances over the negative one Table.\ref{tab:metrics_class}, Fig.\ref{scatter_xgb}. This strongly highlights the advantage of the GWC in producing reliable water spectra, which in turn led to more accurate retrievals. Its success stems from flexible classification, in contrast to the rigid, hard-coded threshold-based approaches used in many algorithms.
	
	A residual analysis of the regression results revealed a systematic error pattern, suggesting that conventional treatment was insufficient and that a more advanced correction method was required. To address this, a three-stage approach was applied, incorporating a CNN-based residual correction model.
	The CNN was trained on normalized residuals between predictions and true values, with an architecture designed to capture both short- and long-range spectral dependencies and detect error structures. Based on the residual analysis, the dataset was split at the 10 mg/m³ threshold, and separate models were trained for each range.
	This approach yielded substantial improvements in absolute error metrics (slope and MAE) and moderate gains in relative error (RMSLE), Table. \ref{log_table},\ref{relres3}, demonstrating the CNN’s effectiveness in mitigating structured residual patterns.
	
	Comparing to relates works, where Sentinel 2 was used to Chl\textit{a} retrieval
	
	The primary limitation of our approach lies in the Sen2Cor atmospheric correction process, which demonstrates relatively poor performance at lower Chl\textit{a} concentrations, which is not fully captured by the ML techniques employed, as evident in the Tables. \ref{tab:cv_entropym}, \ref{tab:cv_entropy_2}. While some of these low values were successfully captured by the CNN-based residual model, Table. \ref{tab:cv_entropy_2}, not all cases were effectively corrected, leaving a gap in sensitivity for this range. Nevertheless, this limitation does not substantially reduce the model’s practical value for policy applications. For water quality policy makers and environmental authorities, the most critical concern lies not in detecting very low Chl\textit{a} levels but in accurately identifying and monitoring medium to high concentrations, particularly those exceeding 30 $mg/m^3$. At these levels, making precise detection is valuable for management strategies.
	
	This framework represents a robust and spatially extensive model that demonstrates strong performance across a wide range of lakes situated in diverse geographical regions. Its adaptability ensures that the methodology can be readily transferred and applied to multiple water bodies, regardless of their location, morphological characteristics, or environmental conditions. Such generalizability underscores the model’s potential as a scalable tool for regional to global monitoring efforts, providing consistent and reliable insights into water quality dynamics across heterogeneous aquatic systems.

Compared to other Chl\textit{a} retrieval studies, our framework exhibits strong generalization without requiring site-specific calibration. While many prior models achieved high local accuracies by optimizing for individual lakes or regions, their transferability often declines when applied to optically distinct systems \cite{joshi2024cross,li2025mlretrieval}. In contrast, our model was trained and validated across diverse inland waters and climatic regimes, yet maintained high predictive performance (\(R^2 = 0.79~(0.75)\), slope \(= 0.91~(0.84)\), MAE \(= 13.52~(15.67)\,\mathrm{mg/m^3}\), and \(\mathrm{RMSLE}=0.56~(0.51)\) in linear-space and log-space, respectively) when evaluated against 867 independent water bodies spanning Chl\textit{a} concentrations from 0 to 1000~\(\mathrm{mg/m^3}\)). Multi-regional Sentinel-2~MSI studies similarly demonstrate strong yet regime-dependent performance: \cite{pahlevan2019seamless} achieved \(\mathrm{RMSLE} < 0.6\) and a slope of approximately \(0.87\) using a Mixture Density Network, \cite{Tran2023_BandRatiosChla} reported a slope of \(0.76\) and \(\mathrm{RMSE} = 0.25\) (\(\mathrm{MAPD} \approx 21.6\%\)) with a coastal multi-regime blend, and \cite{w17111718} obtained \(R^{2} = 0.79\) and \(\mathrm{RMSE} = 10.33~\mathrm{mg/m^3}\) by combining Sentinel-1 and Sentinel-2 data over 35 lakes. Regionally, \cite{lobo2021algaemap} achieved \(R^2 = 0.86\) for the Paraná River basin in Brazil, while \cite{He2022_PSSDFN} reported \(R^2 = 0.77\) and \(\mathrm{RMSE} = 34.77~\mathrm{mg/m^3}\) on a time-consistent test dataset. Collectively, these results emphasize that while Sentinel-based models can achieve high accuracy within specific regimes, globally trained, data-driven frameworks such as ours offer more consistent and transferable Chl\textit{a} predictions across optically diverse inland waters.

\section{ Declaration of competing interests }
	The authors declare that they have no known competing financial
	interests or personal relationships that could have appeared to influence
	the work reported in this paper.
	\section{Appendix}
In the following appendix, for the reader’s convenience, we present a simplified formulation of the algorithm to enable rapid and practical implementation. This reduced approach is intended to facilitate straightforward application and quick evaluation, relying on a limited set of spectral bands to improve computational efficiency while maintaining robust performance, with only a minor increase in error relative to the full model.

Among the tested predictors, the Bloom Index, BI$=R_{705}/R_{665}$ proved to be the most effective descriptor of Chla variability. It exhibited a strong, approximately linear relationship with Chla across the calibrated range, providing a simple yet reliable proxy for rapid bloom assessment and scalable application.
\begin{equation}
	\mathrm{Chl\text{-}a}\,(\mathrm{mg\,m^{-3}}) =
	\begin{cases}
		121.95\,BI - 94.53, & 0.783 < BI \leq 2.0 \\
		175.44\,BI - 201.19, & BI > 2.0
	\end{cases}
\end{equation}

The above formulation reproduces the model output with an average accuracy of approximately 92\% while enabling rapid and straightforward implementation.

	\bibliographystyle{unsrt}
	\bibliography{references}

\end{document}